\documentclass[twocolumn,aps]{revtex4-1}
\usepackage{graphicx}
\usepackage{amsfonts,amssymb}
\usepackage{amsmath}
\usepackage{dcolumn} 
\usepackage{bm}
\usepackage{longtable}
\usepackage{natbib}
\usepackage{xspace}
\usepackage{comment}

\renewcommand{\c}{{\rm c}}

\newcommand{\ci}[0]{c_{I}}

\newcommand{\Space}[0]{\mathcal{S}}
\newcommand{\eref}[0]{E_{\rm CIPSI}}

\newcommand{\psiref}[0]{\Psi_{\rm CIPSI}}

\renewcommand{\b}[1]{\ensuremath{\mathbf{#1}}}

\newcommand{\ket}[1]{\ensuremath{\vert #1  \rangle}}

\newcommand{\bra}[1]{{\ensuremath{\langle #1|}}}
\newcommand{\elemm}[3]{{\ensuremath{\bra{#1}{#2}\ket{#3}}\xspace}}

\newcommand{\ovrlp}[2]{{\ensuremath{\langle #1|#2\rangle}\xspace}}

\newcommand{\kapa}[0]{\ket{\mu}}

\begin{document}

\preprint{}
\title{Quantum Monte Carlo with reoptimized perturbatively selected configuration-interaction wave functions}
\author{Emmanuel Giner$^1$}
\author{Roland Assaraf$^2$}
\author{Julien Toulouse$^2$}
\affiliation{
$^1$Dipartimento di Scienze Chimiche e Farmaceutiche, Universita di Ferrara, Via Fossato di Mortara 17, I-44121 Ferrara, Italy\\
$^2$Sorbonne Universit\'es, UPMC Univ Paris 06, CNRS, Laboratoire de Chimie Th\'eorique, 4 place Jussieu, F-75005, Paris, France}

\date{January 14, 2016}
\begin{abstract}
We explore the use in quantum Monte Carlo (QMC) of trial wave functions consisting of a Jastrow factor multiplied by a truncated configuration-interaction (CI) expansion in Slater determinants obtained from a CI perturbatively selected iteratively (CIPSI) calculation. In the CIPSI algorithm, the CI expansion is iteratively enlarged by selecting the best determinants using perturbation theory, which provides an optimal and automatic way of constructing truncated CI expansions approaching the full CI limit. We perform a systematic study of variational Monte Carlo (VMC) and fixed-node diffusion Monte Carlo (DMC) total energies of first-row atoms from B to Ne with different levels of optimization of the parameters (Jastrow parameters, coefficients of the determinants, and orbital parameters) in these trial wave functions. The results show that the reoptimization of the coefficients of the determinants in VMC (together with the Jastrow factor) leads to an important lowering of both VMC and DMC total energies, and to their monotonic convergence with the number of determinants. In addition, we show that the reoptimization of the orbitals is also important in both VMC and DMC for the Be atom when using a large basis set. These reoptimized Jastrow-CIPSI wave functions appear as promising, systematically improvable trial wave functions for QMC calculations.
\end{abstract}

\maketitle

\section{Introduction}

The development of accurate electronic-structure computational methods remains a topical research subject of high interest. Currently, the two most popular families of electronic-structure computational methods in quantum chemistry are density-functional theory (see, e.g., Ref.~\onlinecite{Koh-RMP-99}) and post-Hartree-Fock wave-function approaches (see, e.g., Ref.~\onlinecite{Pop-RMP-99}). Quantum Monte Carlo (QMC) methods (see, e.g, Refs.~\onlinecite{FouMitNeeRaj-RMP-01,AusZubLes-CR-12,TouAssUmr-INC-15}) are alternative approaches, which become more and more attractive thanks to recent theoretical developments and to the important growing of available computational resources. Indeed, QMC methods need little memory and are embarrassingly parallel which make them ideally suited to modern massively parallel supercomputers.

The two most commonly used variants of QMC methods are variational Monte Carlo (VMC) and fixed-node diffusion Monte Carlo (DMC). The VMC method uses a flexible trial wave function, usually including an explicit correlation factor called the Jastrow factor, and applies Monte Carlo numerical integration techniques for calculating the multidimensional integrals of quantum mechanics. The DMC methods goes beyond VMC by extracting the projection of the trial wave function on the exact ground-state wave function of the system. In practice, except for a few very simple systems, it is necessary to impose the fixed-node approximation in DMC and one only obtains the energy of the best variational wave function having the same nodes as the trial wave function. The construction of optimal trial wave functions is thus crucial for accurate results in both VMC and DMC.

In recent years, a lot of effort has been devoted to developing efficient methods for optimizing a large number of parameters in QMC trial wave functions (see, e.g., Refs.~\onlinecite{FilFah-JCP-00,Sor-PRB-01,UmrFil-PRL-05,Sor-PRB-05,SceFil-PRB-06,TouUmr-JCP-07,UmrTouFilSorHen-PRL-07,TouUmr-JCP-08,BajTiaHooKenReb-PRL-10}). One of the most effective approaches is the linear optimization method of Refs.~\onlinecite{TouUmr-JCP-07,UmrTouFilSorHen-PRL-07,TouUmr-JCP-08}. This is an extension of the zero-variance generalized eigenvalue equation approach of Nightingale and Melik-Alaverdian~\cite{NigMel-PRL-01} to arbitrary nonlinear parameters, and it permits a very efficient and robust energy minimization in a VMC framework. The availability of such optimization methods have recently lead to the exploration of various forms of trial wave functions: Jastrow-antisymmetrized-geminal-power wave functions~\cite{CasSor-JCP-03,CasAttSor-JCP-04,MarAzaCasSor-JCP-09}, Jastrow-pfaffians wave functions~\cite{BajMitDroWag-PRL-06,BajMitWagSch-PRB-08}, Jastrow-backflow wave functions~\cite{LopMaDruTowNee-PRE-06}, orbital-attached multi-Jastrow wave functions~\cite{BouBraCaf-JCP-10}, and various types of Jastrow-valence-bond wave functions~\cite{AndGod-JCP-10,BraTouCafUmr-JCP-11,FraFilAmo-JCTC-12}.

For atoms and molecules, the most used and systematically improvable form of QMC trial wave functions remains a Jastrow factor multiplied by a truncated configuration-interaction (CI) expansion in Slater determinants (see, e.g., Refs.~\onlinecite{FilUmr-JCP-96,TouUmr-JCP-07,UmrTouFilSorHen-PRL-07,BroTraLopNee-JCP-07,TouUmr-JCP-08,ZimTouZhaMusUmr-JCP-09,BajTiaHooKenReb-PRL-10,needs_atoms,TouCafReiHogUmr-INC-12,PetTouUmr-JCP-12,MorMcmClaKimScu-JCTC-12}). However, a major problem is how to systematically select the best determinants entering the truncated CI expansion. Standard CI calculations are usually truncated based on orbital active space and/or excitation criteria, which are far from being optimal for rapidly approaching the full CI (FCI) limit.

Very recently, Giner \textit{et al.}~\cite{canadian,atoms_3d,giner_phd,f2_dmc} have proposed to use in QMC truncated CI expansions constructed from the CI perturbatively selected iteratively (CIPSI) algorithm~\cite{cipsi2}. In this approach, the truncated CI expansion is iteratively enlarged by selecting the most important missing determinants using perturbation theory. The advantage of such an approach resides in the fact that the selected determinants are by construction those having the largest impact on electronic correlation, both of static and dynamic nature, in the specific system under consideration. This algorithm thus allows one to systematically and rapidly approach the FCI wave function in an automatic way, and obtain truncated CI expansions with limited numbers of determinants that are quantitative approximations to the FCI wave function. Giner \textit{et al.} used these CIPSI wave functions in DMC calculations, without any Jastrow factor and without reoptimizing any parameters in QMC, and showed that, with enough determinants, accurate results can be obtained with these trial wave functions for atomic and molecular systems. 

The present work continues this line of research by studying the effect of adding a sophisticated Jastrow factor to these CIPSI wave functions and optimizing the parameters (Jastrow parameters, coefficients of the determinants, and orbital parameters) in VMC using the linear method. In particular, we want to know whether the reoptimization in VMC of the coefficients of the determinants and/or the orbitals in these trial wave functions in the presence of the Jastrow factor brings important improvements in VMC and DMC. The paper is organized as follows. Section \ref{theory} describes the methodology of the CIPSI algorithm, the parametrization of the Jastrow-CIPSI wave functions used in QMC, and gives computational details on the calculations performed. Section \ref{results} gives and discusses the numerical results obtained on first-row atoms from Be to Ne. Finally, we give our conclusions in Section \ref{conclusions}.

\section{Methodology}
\label{theory}

\subsection{CIPSI wave functions}

The CIPSI method is a selected CI algorithm where the determinants are chosen according to a perturbative estimation of their importance, which allows one to build CI expansions keeping only the most important determinants. As this idea is somewhat intuitive, various selected CI schemes guided by perturbation theory have been proposed \cite{bender,malrieu,buenker1,buenker-book,cipsi2,harrison}, but the CIPSI version is one of the best algorithms in this field, as it introduces an iterative procedure and various stopping criteria to control the quality of the wave functions and energies. The convergence of the CIPSI wave functions and energies has been intensively studied in the past decades \cite{cipsi2,Rubio198698,cimiraglia_cipsi,cele_cipsi_zeroth_order,Angeli2000472} and a revival of such ideas has appeared in the DMC framework in the last few years \cite{canadian,atoms_3d,f2_dmc}. We now summarize the CIPSI algorithm used in this work.

At the beginning of a given CIPSI iteration, one has a reference CIPSI wave function $\ket{\Psi_{\rm CIPSI}}$ built with Slater determinants $\ket{I}$ that span a space $\mathcal{S}$,
\begin{equation} 
\label{cipsi1}
\ket{\Psi_{\rm CIPSI}} = \sum_{I \in \mathcal{S}} c_{I} \ket{I},
\end{equation}
and the coefficients $c_{I}$ are obtained by minimization of the variational energy $E^{(0)}$,
\begin{equation} 
E^{(0)} = \min_{\{ c_{I} \}} \frac{\elemm{\Psi_{\rm CIPSI}}{\hat{H}}{\Psi_{\rm CISPI}}}{\ovrlp{\Psi_{\rm CIPSI}}{\Psi_{\rm CIPSI}}},
\end{equation}
where $\hat{H}$ is the many-body Hamiltonian operator. One then considers the determinants that do not belong to the $\mathcal{S}$ space. For each such determinant $\kapa$, its first-order coefficient using the Epstein-Nesbet zeroth-order Hamiltonian \cite{epstein,nesbet} is given by
\begin{equation} 
\begin{aligned}
 c_{\mu}^{(1)} & = \frac{\elemm{\mu}{\hat{H}}{\psiref}}{\eref - \elemm{\mu}{\hat{H}}{\mu}} \\ 
                 & = \sum_{I \in \Space} \ci \,\, \frac{\elemm{\mu}{\hat{H}}{I}}{\eref - \elemm{\mu}{\hat{H}}{\mu}}, 
\end{aligned}
\end{equation}
and then the contribution of $\kapa$ at second order to the energy is
\begin{equation} 
\begin{aligned}
 e_{\mu}^{(2)} & = c_{\mu}^{(1)} \elemm{\psiref}{\hat{H}}{\mu} \\ 
                 & = \frac{|\elemm{\mu}{\hat{H}}{\psiref}|^2}{\eref - \elemm{\mu}{\hat{H}}{\mu}}.
\end{aligned}
\end{equation}
At each CIPSI iteration, the total second-order correction to the energy $E^{(2)}$ is defined as the sum of all the energy contributions $e_{\mu}^{(2)}$ over all $\ket{\mu}$ not belonging to the $\mathcal S$ space,
\begin{equation} 
\label{ept2}
\begin{aligned}
 E^{(2)} & = \sum_{\mu \notin \mathcal S } e_{\mu}^{(2)},
\end{aligned}
\end{equation}
and the CIPSI energy is defined as $E_{\rm CIPSI} = E^{(0)} + E^{(2)}$, which is an estimation of the FCI energy.
In the version of CIPSI used here, the determinant $\kapa$ is added to the $\mathcal{S}$ space if its second-order contribution to the energy $e_{\mu}^{(2)}$ is larger than a certain threshold $\eta$ set at the beginning of the iteration. When the selection procedure has been done, one obtains a new CIPSI wave function constructed with a new set determinants belonging to the new $\mathcal{S}$ space (i.e., the determinants present at the beginning of the iteration together with the ones that have just been added). This new CIPSI wave function serves as the reference wave function for the next iteration. The selection threshold $\eta$ is lowered at each iteration, and this process is repeated until one reaches a given criterion of convergence. The criterion used here is a given threshold on the remaining second-order contribution to the energy $E^{(2)}$~\cite{f2_dmc}, but one can find in the literature other stopping criteria based on an estimation of the norm of the first-order wave function~\cite{cele_cipsi_zeroth_order} or simply based on the number of determinants belonging to the $\mathcal{S}$ space. It should be noted that thanks to perturbation theory and to its iterative nature, the CIPSI algorithm allows one to select determinants in the whole FCI space, whereas standard truncated CI methods restrict the determinants to a given pattern of excitations with respect to a reference space and/or an orbital space. 

\subsection{Jastrow-CIPSI wave functions}

After the CIPSI calculation is done, we construct a Jastrow-CIPSI wave function parametrized as
\begin{equation} 
\label{cipsi1}
\ket{\Psi_{\rm JCIPSI}(\b{p})} = \hat{J}(\bm{\alpha}) \; e^{\hat{\kappa}(\bm{\kappa})} \; \sum_{I \in \mathcal{S}} c_{I} \ket{I},
\end{equation}
where $\hat{J}(\bm{\alpha})$ is a Jastrow-factor operator depending on some parameters $\bm{\alpha}$, and $e^{\hat{\kappa}(\bm{\kappa})}$ is an orbital rotation operator. 

The orbital excitation operator is defined as $\hat{\kappa}(\bm{\kappa}) =\sum_{k<l} \kappa_{kl}(\hat{E}_{kl} - \hat{E}_{lk})$ where $\kappa_{kl}$ are the orbital rotation parameters and $\hat{E}_{kl}=\hat{a}_{k\uparrow}^\dagger \hat{a}_{l\uparrow} + \hat{a}_{k\downarrow}^\dagger \hat{a}_{l\downarrow}$ is the spin-singlet excitation operator from orbital $l$ to orbital $k$. The use of the unitary operator $e^{\hat{\kappa}(\bm{\kappa})}$ allows one to have a non-redundant parametrization of the orbital coefficients, automatically preserving the orthonormality of the orbitals (see Refs.~\onlinecite{TouUmr-JCP-07,TouUmr-JCP-08}). The orbitals are partitioned into three sets: inactive (doubly occupied in all determinants), active (occupied in some determinants and unoccupied in others), and virtual (unoccupied in all determinants). The non-redundant excitations to consider \textit{a priori} are: inactive $\to$ active, inactive $\to$ virtual, active $\to$ virtual, and active $\to$ active. Some redundancies can actually occur with the active-active excitations, and they must be detected and eliminated. Also, only excitations between orbitals of the same irreducible representation need to be considered.

We use a flexible Jastrow factor consisting of the exponential of the sum of electron-nucleus, electron-electron, and electron-electron-nucleus terms, written as systematic polynomial and Pad\'e expansions~\cite{Umr-UNP-XX} (see, also, Refs.~\onlinecite{FilUmr-JCP-96,GucSanUmrJai-PRB-05}) with 24 free parameters. The total parameters $\b{p}=(\bm{\alpha},\b{c},\bm{\kappa})$ to be optimized in the wave function are the Jastrow parameters $\bm{\alpha}$, the coefficients of the determinants $\b{c}$, and the orbital rotation parameters $\bm{\kappa}$.

\subsection{Computational details}
\label{comp_details}

In the present work, the starting orbitals, both occupied and unoccupied, are obtained from restricted Hartree-Fock (RHF) or restricted open-shell Hartree-Fock (ROHF) calculations performed using the GAMESS(US) package \cite{gamess}. 
The basis set used here for the CIPSI calculations is a fit of the polarized triple-zeta VB1 Slater basis set of Ref.~\onlinecite{EmaGarRamLopFerMeiPal-JCC-03}. For the case of the Be atom, we also perform calculations with the polarized quadruple-zeta VB2 Slater basis set.
Each Slater basis function is fitted to a linear combination of 10 Gaussian basis functions~\cite{HehStePop-JCP-69,Ste-JCP-70,KolReiAss-JJJ-XX}. The CIPSI calculations are performed using the Quantum Package~\cite{quantum_package}. The 1s electrons are kept frozen in all calculations, consistently with the fact that the VB1 or VB2 basis set does not provide any basis functions adapted for core-core or core-valence correlation. For all the systems, the starting coefficients of the Slater determinants are obtained from a large CIPSI calculation where the remaining second-order energy correction $|E^{(2)}|$ [see Eq.~(\ref{ept2})] is smaller than $10^{-4}$ hartree. In practice, for the VB1 basis set, the number of determinants in the large CIPSI calculations ranges from 19 for the Be atom to approximatively $10^4$ for the Ne atom. Then, the determinants are sorted according to the absolute values of their coefficients, and the wave function is truncated at various numbers of determinants keeping the coefficients as they are in the large CIPSI wave function (i.e., at the near FCI level). We prefer to use this procedure instead of using the coefficients from CIPSI wave functions with small numbers of determinants. Indeed, the selected determinants and their coefficients from a near FCI calculation are expected to be more optimal for VMC in the presence of the Jastrow factor or for DMC. Nevertheless, for the best possible results in VMC, the selection of the determinants should be done in VMC in the presence of the Jastrow factor, which is not currently implemented.

These CIPSI wave functions are then multiplied by our Jastrow factor, and QMC calculations are performed with the program CHAMP~\cite{Cha-PROG-XX} using the true Slater basis set rather than its Gaussian expansion. The parameters are optimized by minimizing the energy with the linear optimization method~\cite{TouUmr-JCP-07,UmrTouFilSorHen-PRL-07,TouUmr-JCP-08} in VMC, using an accelerated Metropolis algorithm~\cite{Umr-PRL-93,Umr-INC-99}. We test three levels of optimization: optimization of the Jastrow parameters only, simultaneous optimization of the Jastrow parameters and the coefficients of the determinants, and simultaneous optimization of the Jastrow parameters, the coefficients of the determinants, and the orbital parameters. Once the trial wave functions have been optimized, we perform DMC calculations within the short-time and fixed-node approximations (see, e.g., Refs.~\onlinecite{GriSto-JCP-71,And-JCP-75,And-JCP-76,ReyCepAldLes-JCP-82,MosSchLeeKal-JCP-82}). We use an imaginary time step of $\tau=0.0025$ hartree$^{-1}$ in an efficient DMC algorithm with very small time-step errors~\cite{UmrNigRun-JCP-93}.

\begin{figure*}
\includegraphics[scale=0.30,angle=-90]{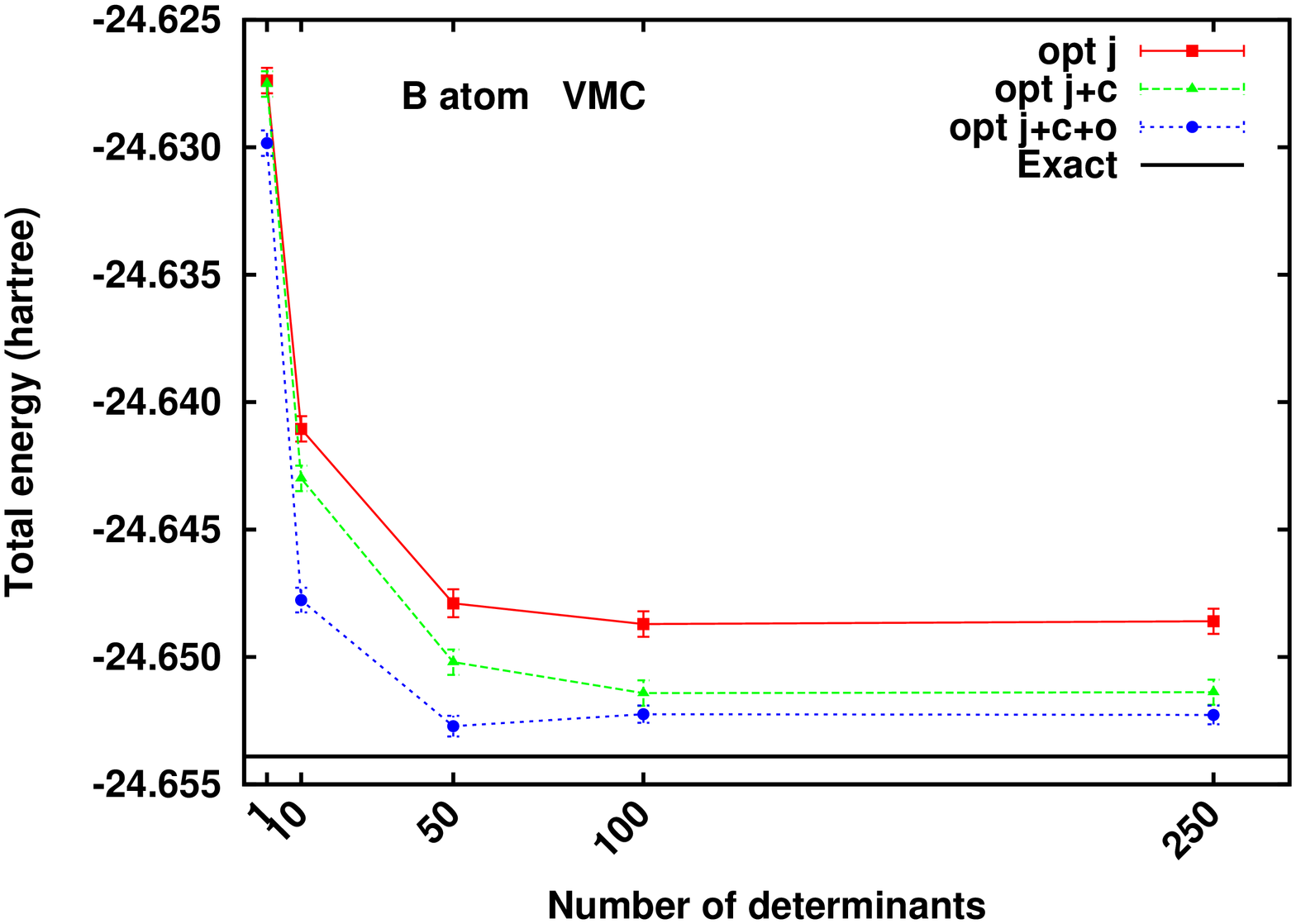}
\includegraphics[scale=0.30,angle=-90]{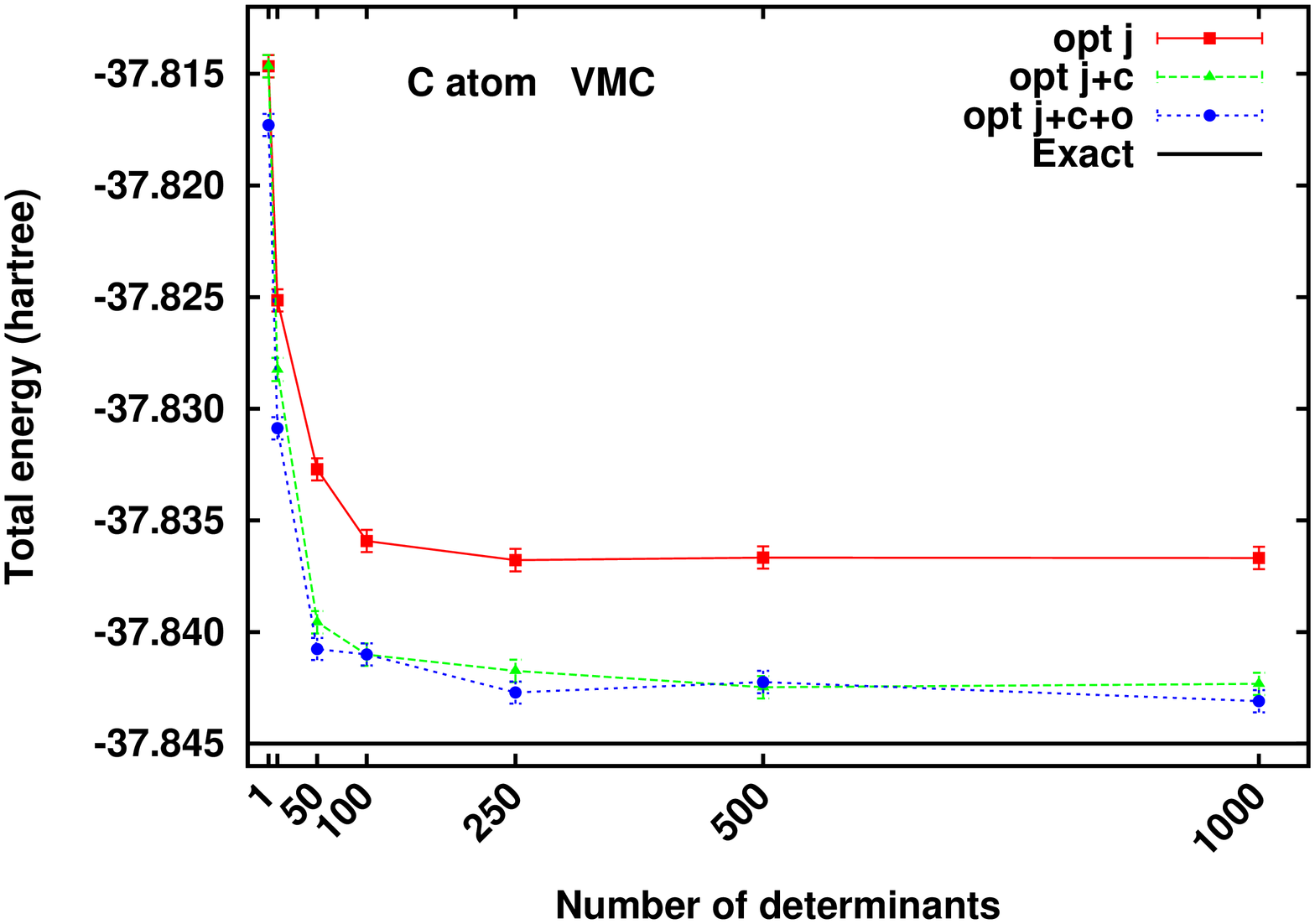}
\includegraphics[scale=0.30,angle=-90]{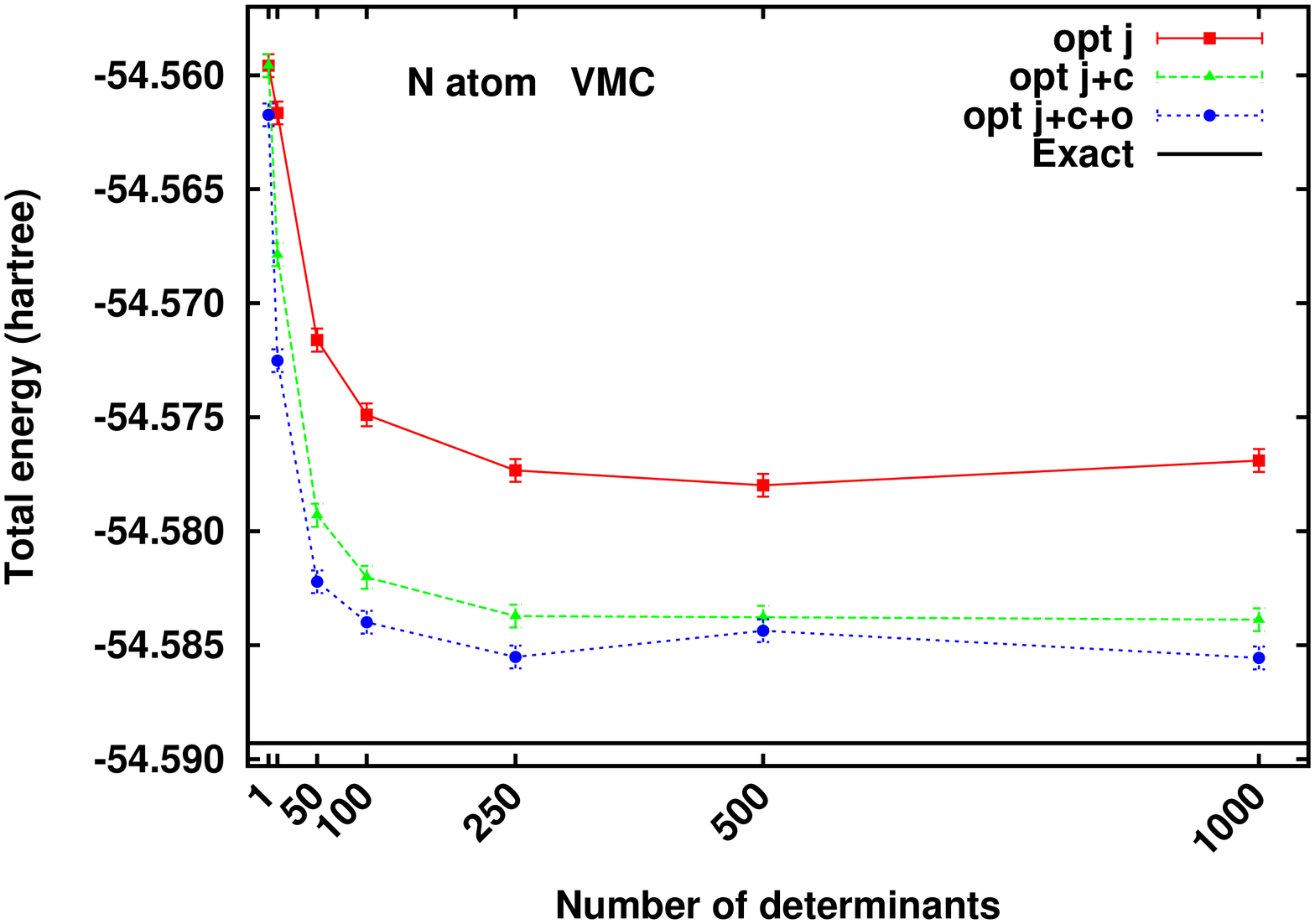}
\includegraphics[scale=0.30,angle=-90]{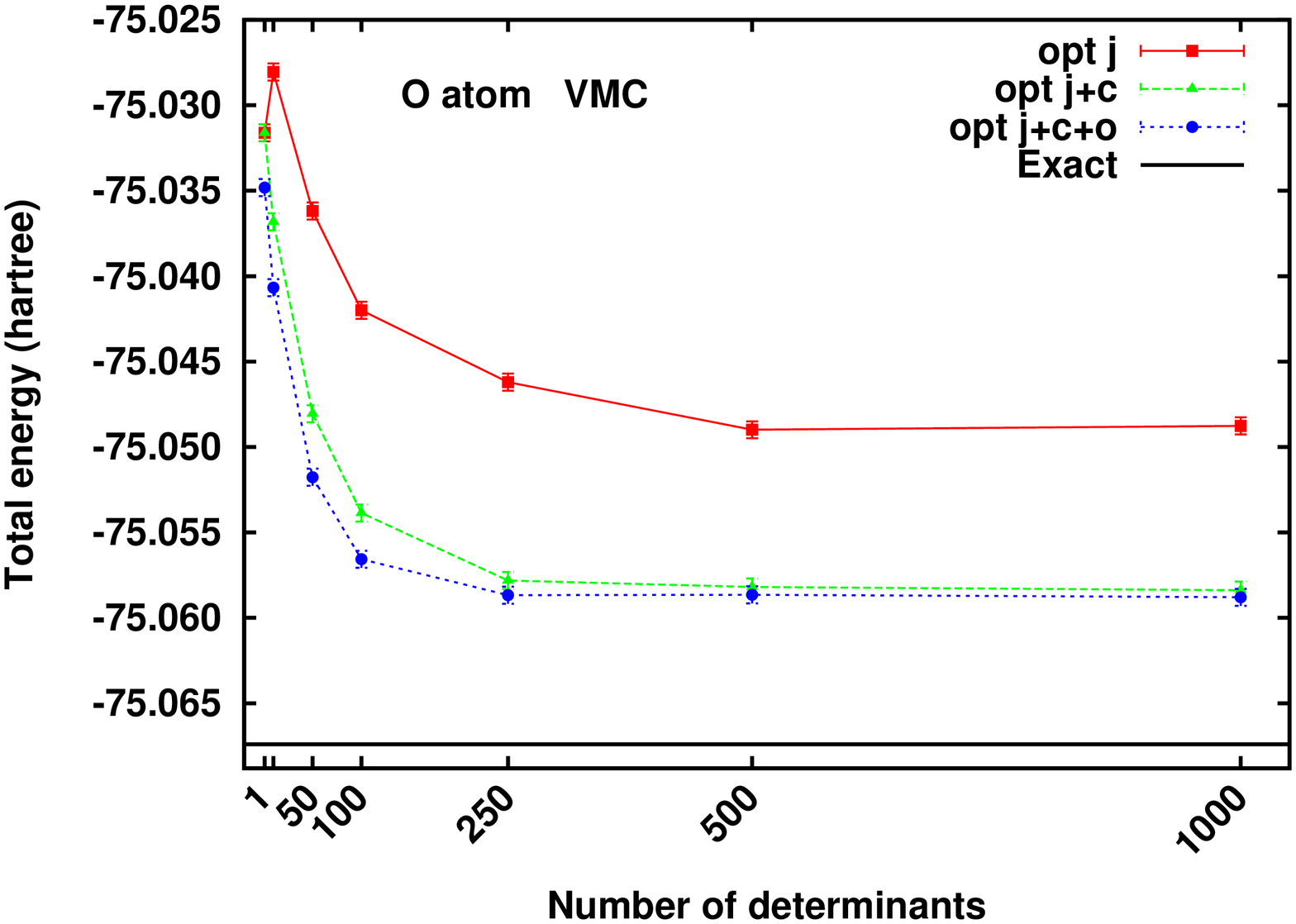}
\includegraphics[scale=0.30,angle=-90]{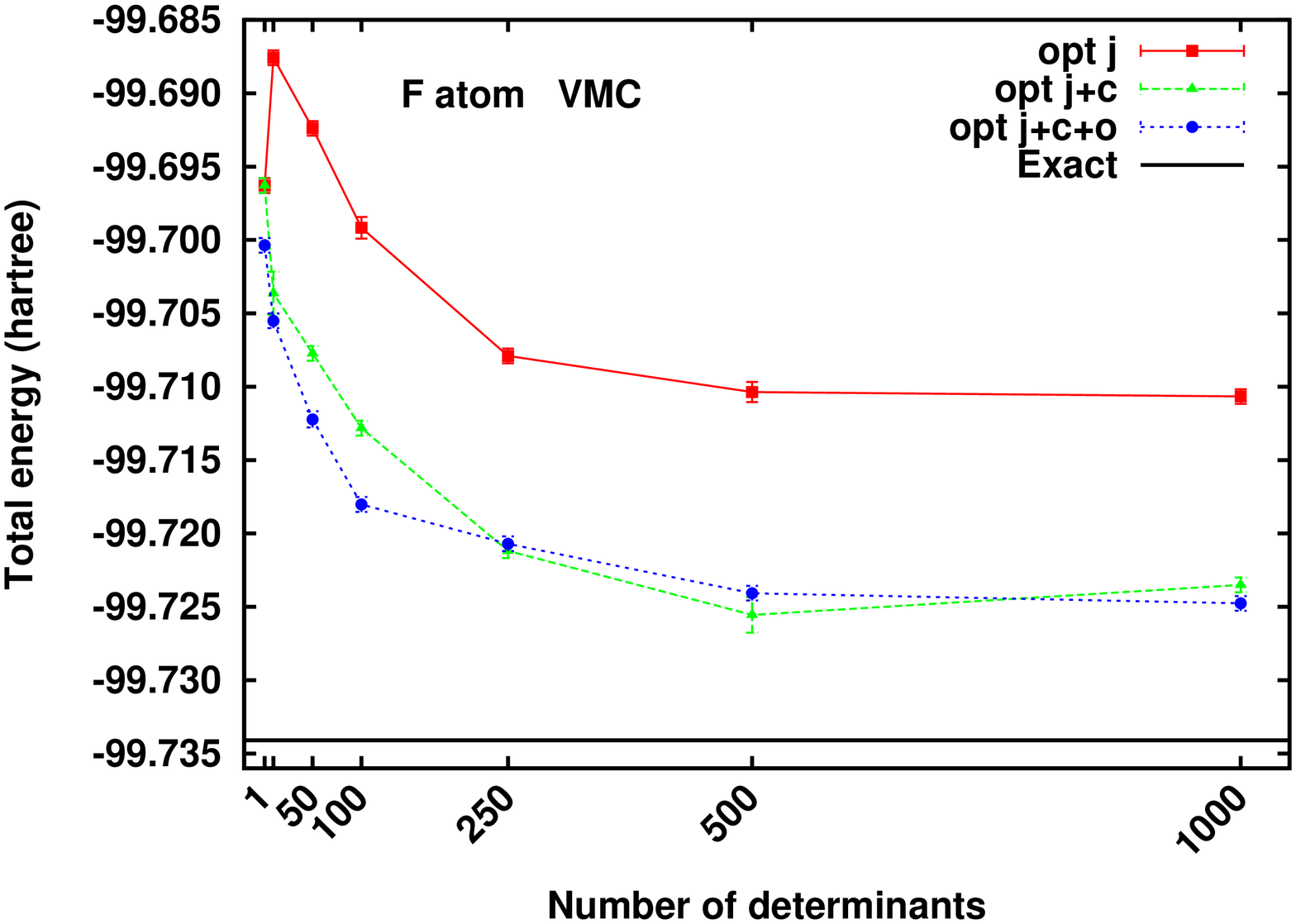}
\includegraphics[scale=0.30,angle=-90]{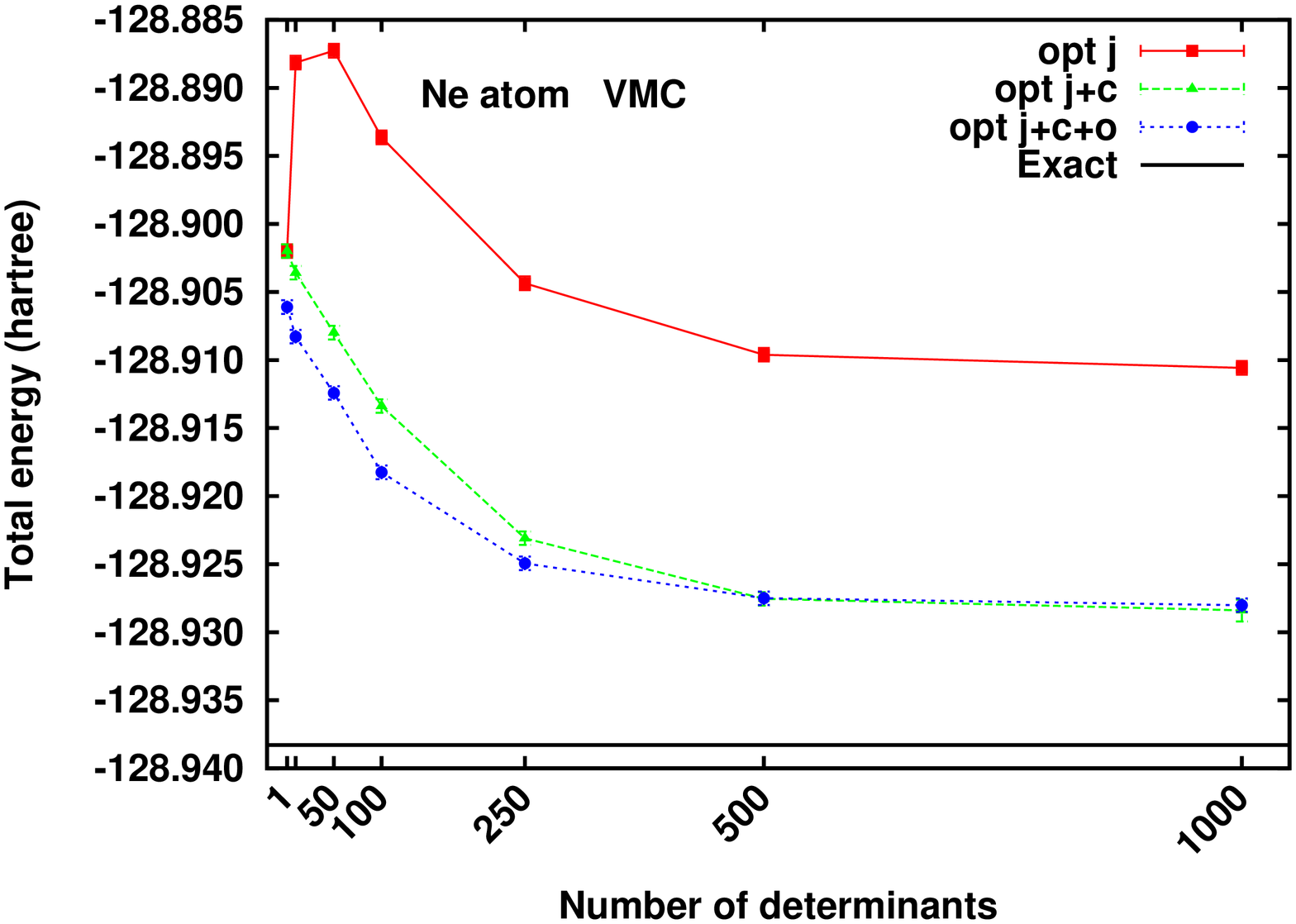}
\caption{VMC total energies of atoms with Jastrow-CIPSI wave functions with increasing numbers of determinants and different levels of optimization in VMC: optimization of the Jastrow factor (opt j), optimization of the Jastrow factor and the coefficients of the determinants (opt j+c), and optimization of the Jastrow factor, the coefficients of the determinants, and the orbitals (opt j+c+o). The basis set used is VB1.
}
\label{conv_neutral_vmc}
\end{figure*}

\begin{figure*}
\includegraphics[scale=0.30,angle=-90]{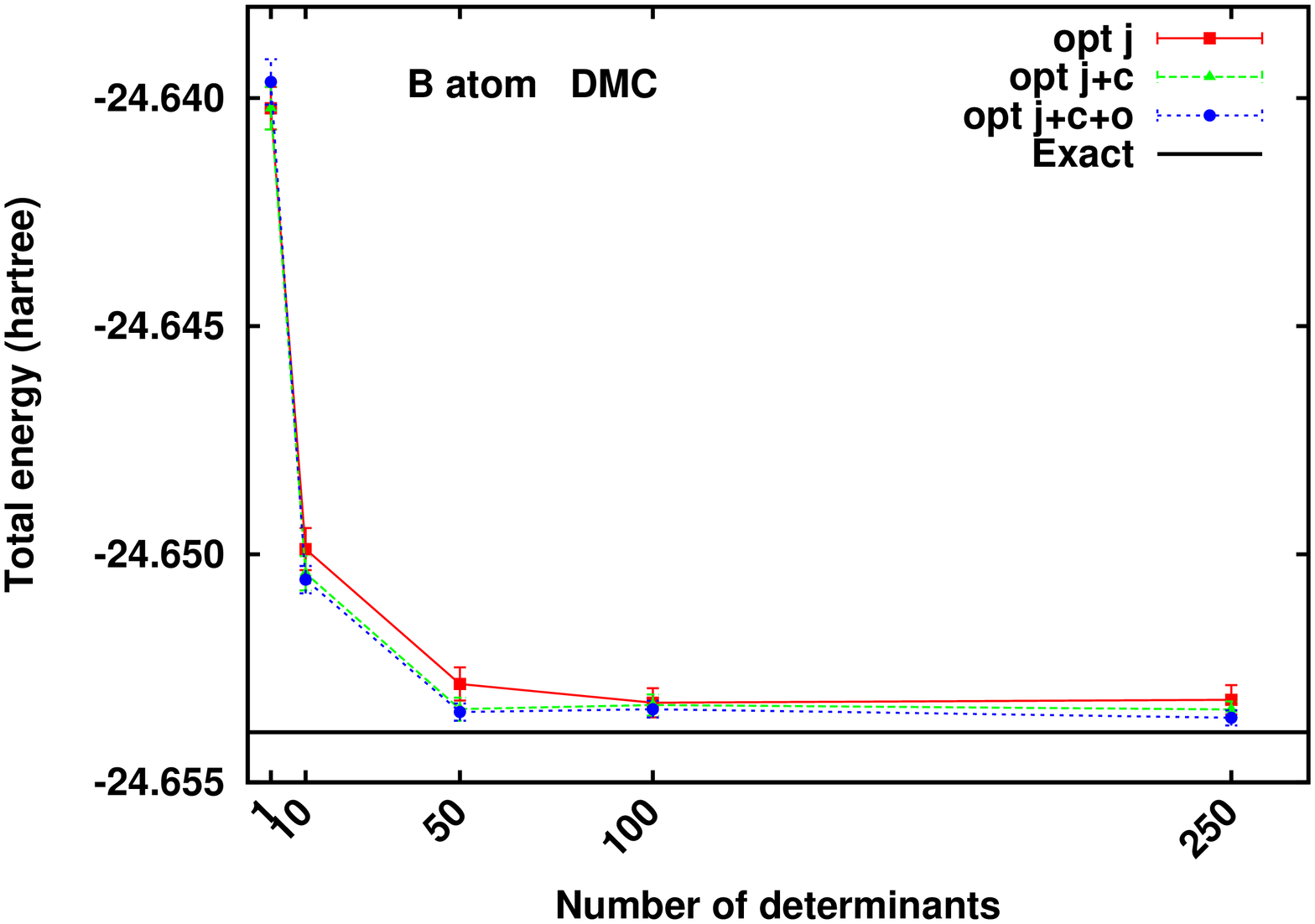}
\includegraphics[scale=0.30,angle=-90]{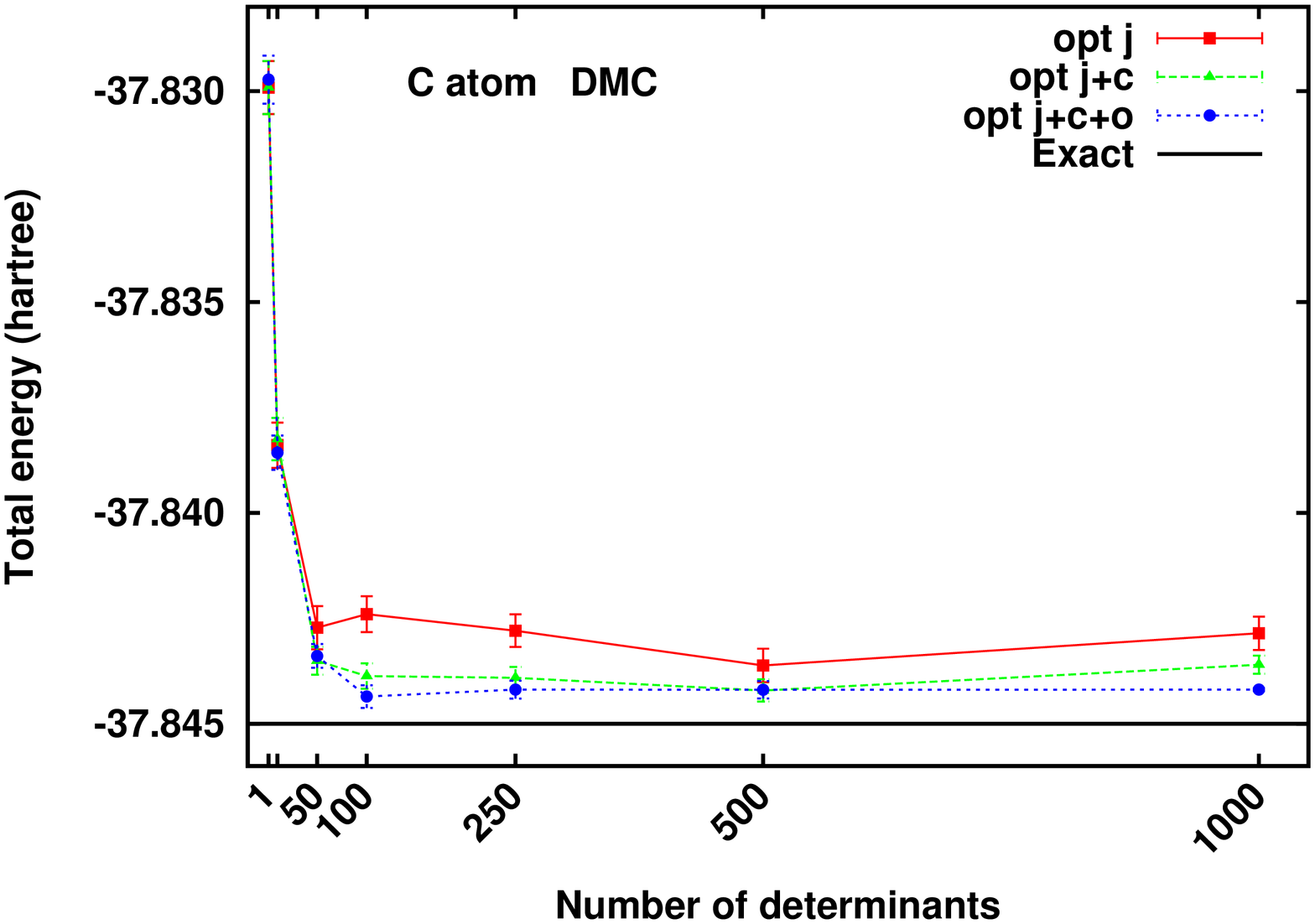}
\includegraphics[scale=0.30,angle=-90]{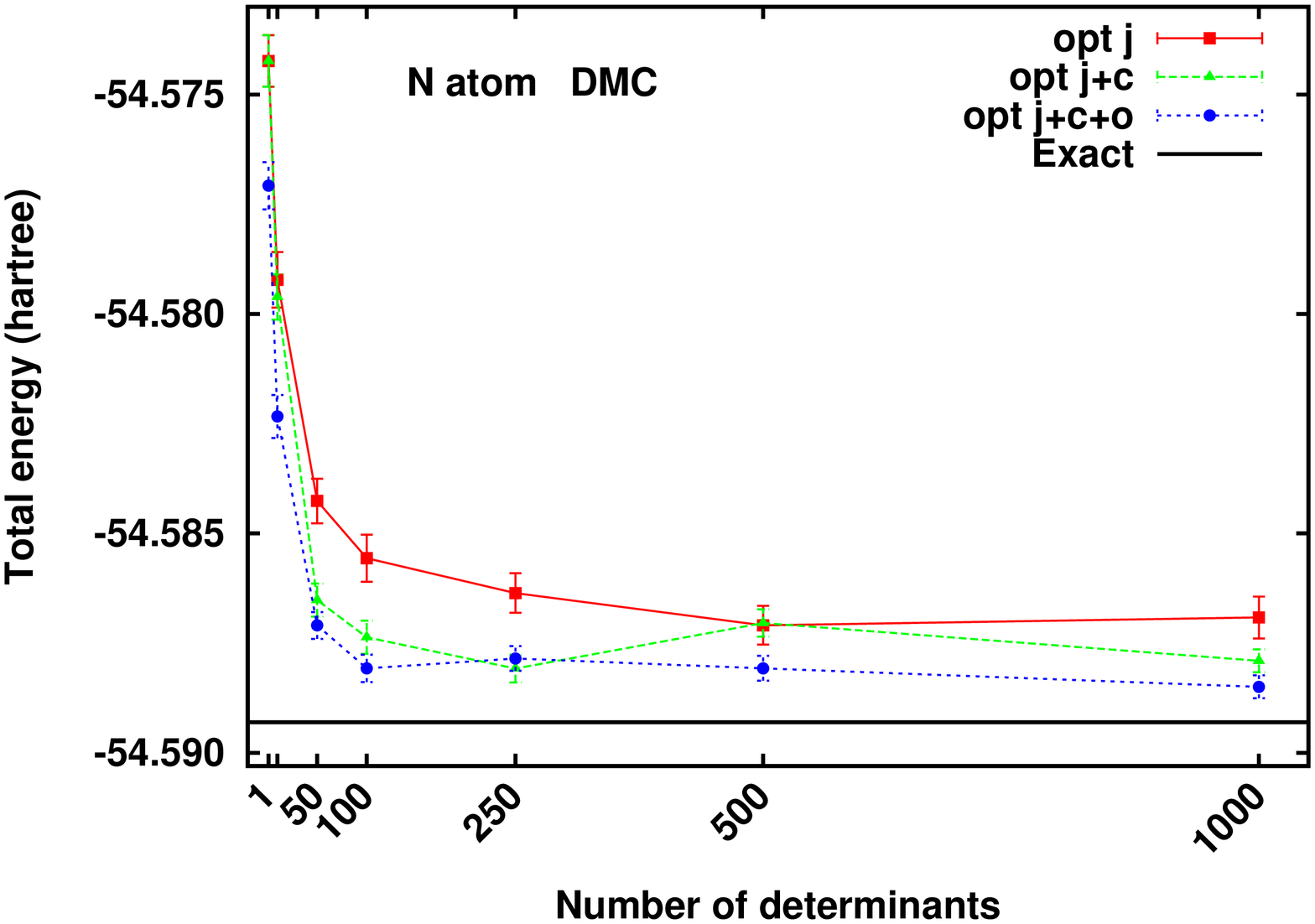}
\includegraphics[scale=0.30,angle=-90]{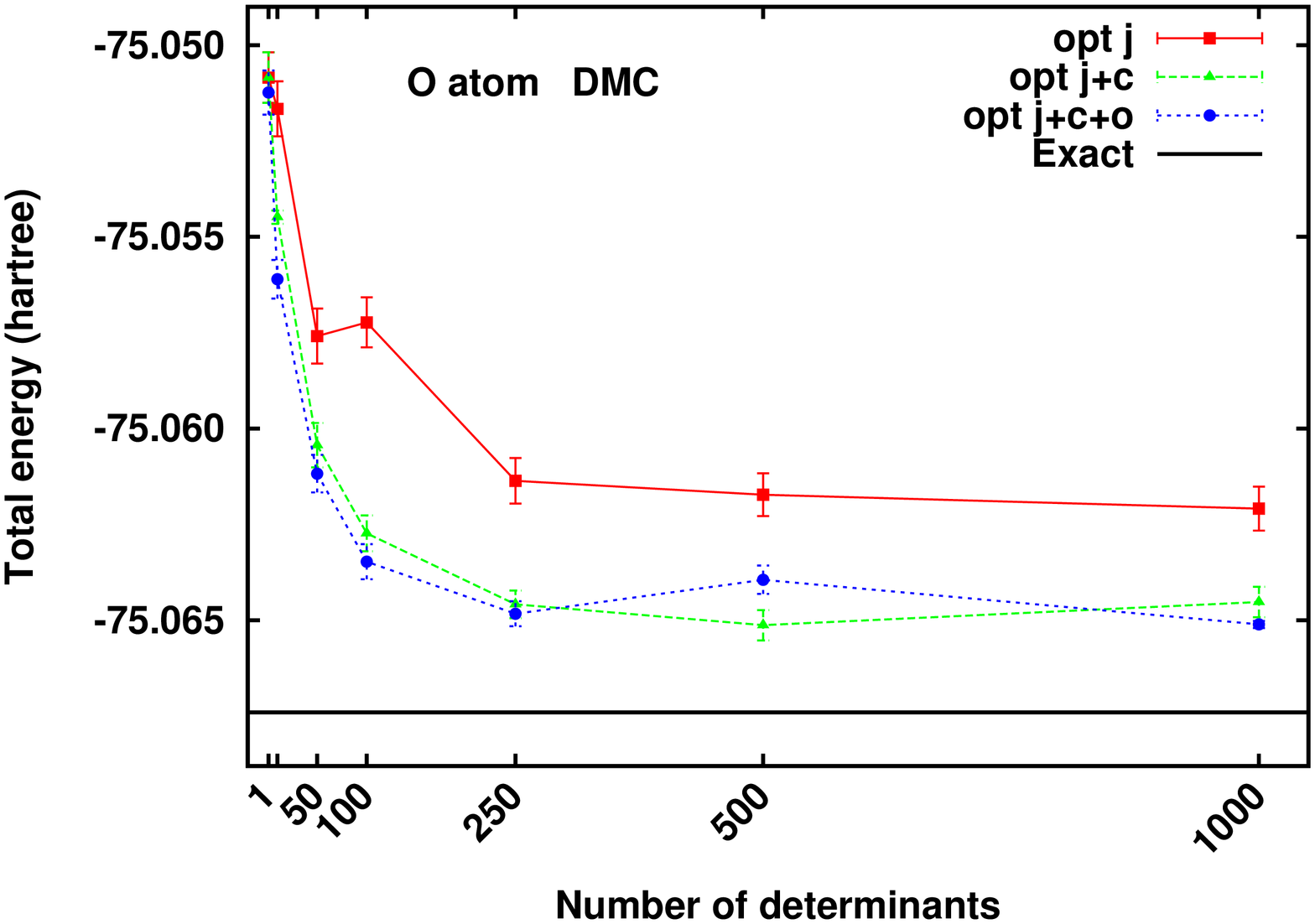}
\includegraphics[scale=0.30,angle=-90]{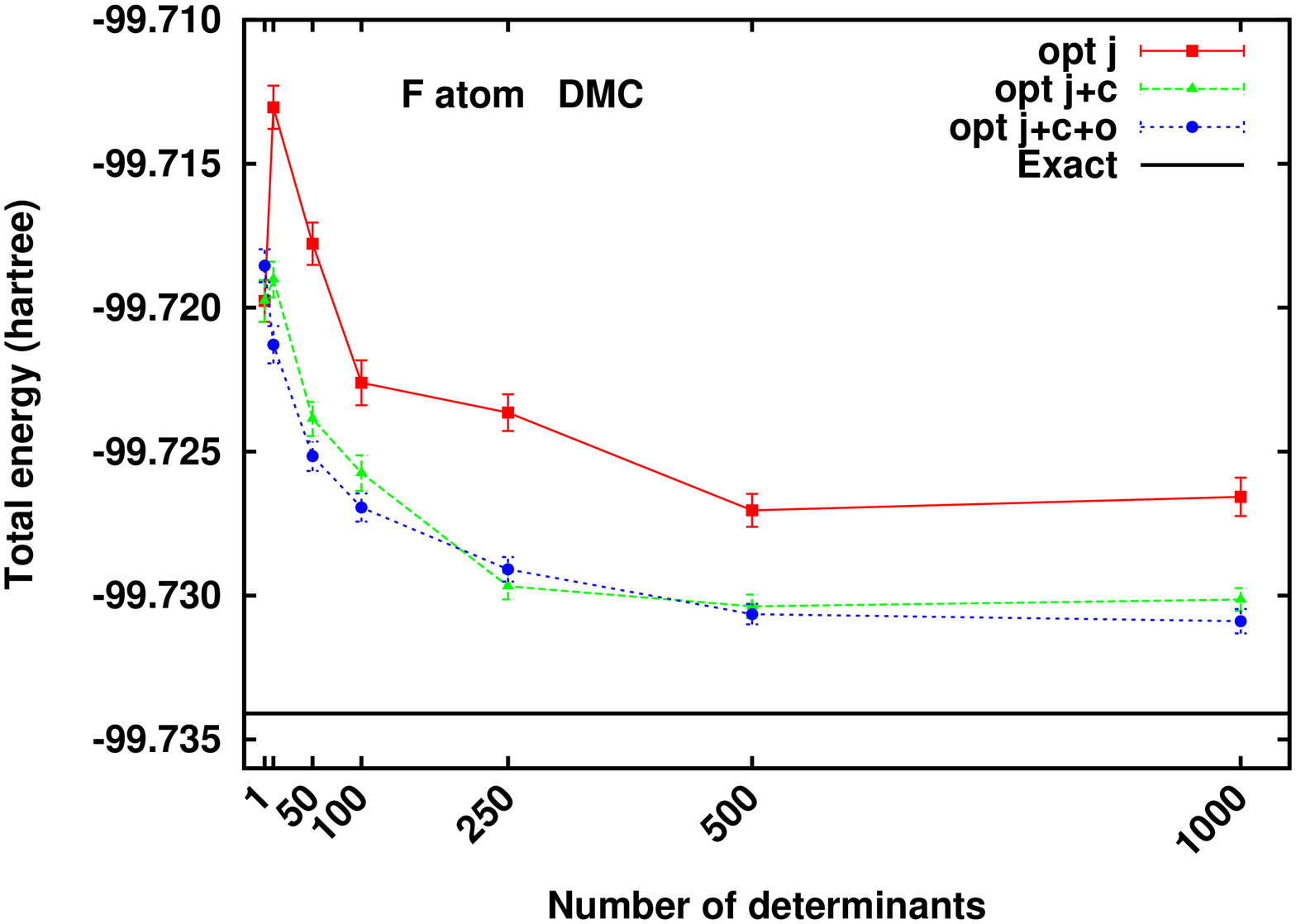}
\includegraphics[scale=0.30,angle=-90]{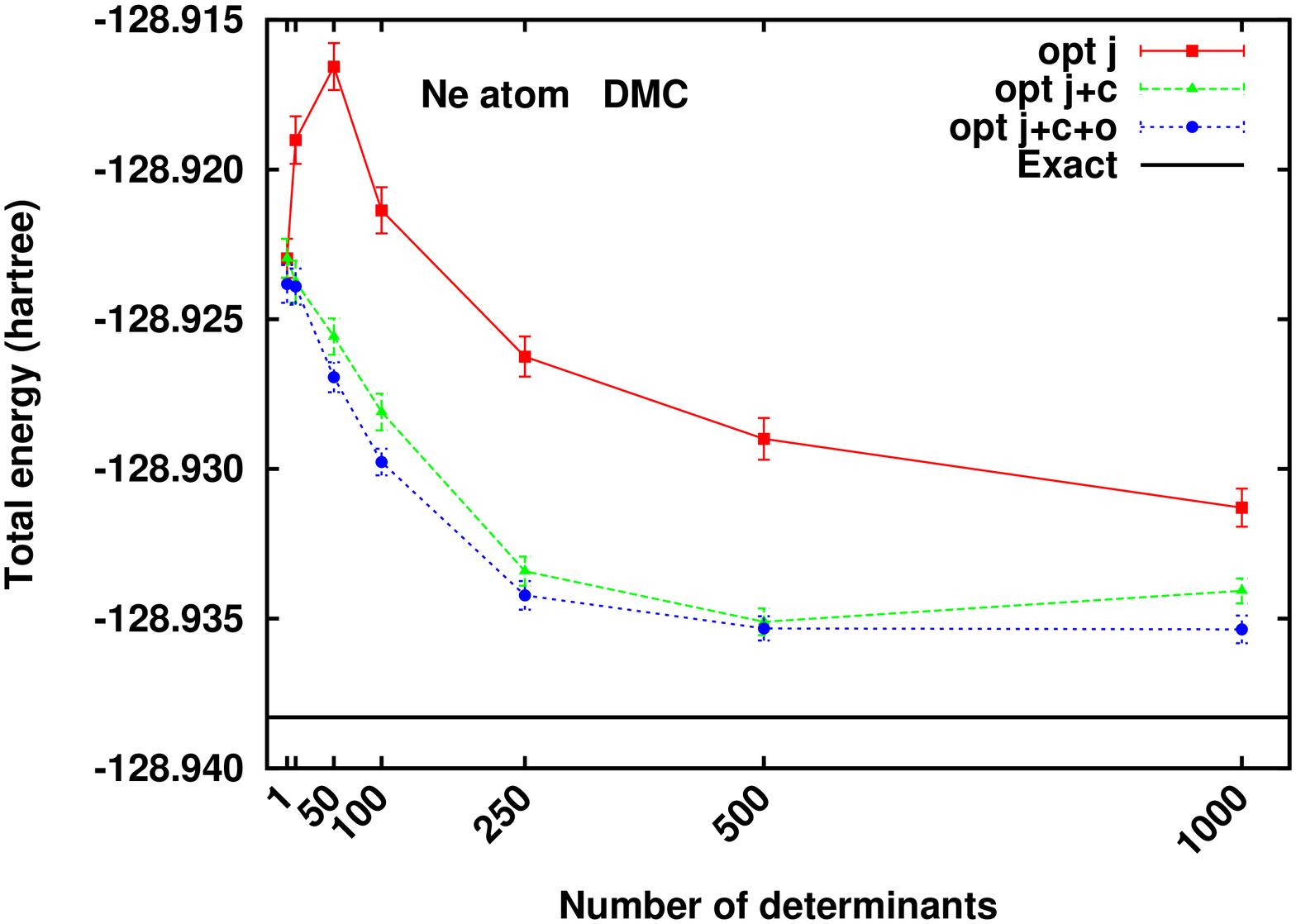}
\caption{DMC total energies of atoms with Jastrow-CIPSI wave functions with increasing numbers of determinants and different levels of optimization in VMC: optimization of the Jastrow factor (opt j), optimization of the Jastrow factor and the coefficients of the determinants (opt j+c), and optimization of the Jastrow factor, the coefficients of the determinants, and the orbitals (opt j+c+o). The basis set used is VB1.
}
\label{conv_neutral_dmc}
\end{figure*}

\section{Results and discussion}
\label{results}

Section \ref{results_b_ne} reports the numerical results obtained for the series of atoms ranging from B to Ne. Then, the case of the Be atom is investigated in more details in Section \ref{results_be} as a prototype of a system showing important static correlation effects. 

\subsection{Results for atoms from B to Ne}
\label{results_b_ne}

In Figure \ref{conv_neutral_vmc} we report the convergence of the total VMC energies of the atoms with respect to the number of determinants at various optimization levels. As expected at the VMC level, for a given number of determinants in the CI expansion, the energy lowers (within the statistical uncertainty) as one increases the number of optimized variational parameters in the wave function. Considering the convergence of the VMC energy as a function of the number of determinants, several observations can be made. First, optimizing only the Jastrow factor, and thus keeping the CI coefficients optimized at the near FCI level, does not lead to a systematically monotonic lowering of the VMC energy upon increasing the number of determinants (cases of the O, F, and Ne atoms). Nevertheless, after reaching a certain number of determinants, the VMC energy tends to lower and converge for a large number of determinants. Second, the optimization of the CI coefficients in the presence of the Jastrow factor removes this non-monotonic behavior of the VMC energy, as it should. Also, the gain in the total VMC energy is important in all the calculations reported here, including the ones with the largest numbers of determinants. Third, the gain in the total VMC energy brought by the optimization of the orbitals (together with the Jastrow factor and the CI coefficients) is much smaller and tends to reduce when increasing the number of determinants. This is expected since in the FCI limit the wave function becomes invariant with respect to orbital rotations.

\begin{table*}[t]
\caption{Total VMC and DMC energies of atoms with Jastrow-CIPSI wave functions with the largest numbers of determinants used ($N_\text{det}^\text{max}$) and different levels of optimization in VMC: optimization of the Jastrow factor (opt j), optimization of the Jastrow factor and the coefficients of the determinants (opt j+c), and optimization of the Jastrow factor, the coefficients of the determinants, and the orbitals (opt j+c+o). The basis set used is VB1.}
\label{final_energies}
\begin{tabular}{l l c c c c c c c c c}
\hline\hline
        &       & \multicolumn{3}{c}{$E_\text{VMC}$} &&& \multicolumn{3}{c}{$E_\text{DMC}$} &  $E_\text{exact}^a$\\
                \cline{3-5} \cline{7-10} 
        & $N_\text{det}^\text{max}$  & opt j          & opt j+c          & opt j+c+o        &&& opt j           & opt j+c         & opt j+c+o      &   \\
\hline
B       & 250                 & -24.6486(5) & -24.6514(5)  &  -24.6523(4) &&& -24.6532(3) & -24.6534(2) & -24.6536(2) & -24.65390   \\
C       & 1000            & -37.8367(5) & -37.8423(5)  & -37.8431(5)  &&& -37.8429(4) & -37.8436(2) & -37.8442(1)  & -37.8450 \\
N       & 1000            & -54.5769(5) & -54.5839(5)  & -54.5856(5)  &&& -54.5869(5) & -54.5879(3) & -54.5885(3) & -54.5893\\
O       & 1000            & -75.0488(5) & -75.0584(5)  & -75.0588(5)  &&& -75.0621(6) & -75.0645(4) & -75.0651(1)  & -75.0674\\
F       & 1000            & -99.7107(5) & -99.7235(5)  & -99.7248(5)  &&& -99.7266(7) & -99.7301(4) & -99.7309(4) & -99.7341\\
Ne      & 1000            & -128.9106(5) & -128.9284(8)& -128.9280(5) &&& -128.9313(6) & -128.9341(4)& -128.9354(5)& -128.9383 \\
\hline\hline
\multicolumn{11}{l}{$^a$Estimated non-relativistic total energies from Ref.~\onlinecite{davidson}.}
\end{tabular}
\end{table*}

Figure~\ref{conv_neutral_dmc} shows the convergence of the DMC total energies as a function of the number of determinants using the Jastrow-CIPSI wave functions previously optimized at the VMC level. Similarly to what was observed for the VMC total energies, when only the Jastrow factor has been optimized, the DMC total energies do not systematically converge monotonically with the number of determinants. A transient region clearly occurs for the F and Ne atoms where the DMC energy increases with respect to the one obtained using the RHF/ROHF determinant. For these atoms, obtaining a DMC energy lower than the one obtained with the RHF/ROHF determinant requires at least 100 and 250 determinants, respectively. For larger numbers of determinants, the DMC energy with the non-reoptimized CI expansions does tend to decrease monotonically with the number of determinants. Moving now to the DMC results using the wave functions where the Jastrow factor and the CI coefficients have been simultaneously optimized, two observations can be made. First, at a given number of determinants, a substantially lower DMC energy is obtained in comparison to the one obtained without the reoptimization of the CI coefficients (except for the B atom, for which the gain is comparable to the statistical uncertainty). Thus, the CI coefficients obtained from the large CIPSI wave function clearly do not provide the best nodal structure, and the dynamical correlation brought by the Jastrow factor significantly changes the CI coefficients and improves the nodes of the wave function. Second, the previously observed non-monotonic behavior of the DMC energy at small numbers of determinants is avoided, and the DMC energy now converges monotonically (within the statistical uncertainties) and more rapidly with the number of determinants. Even though having a monotonic convergence seems reasonable, note that there is in principle no guarantee that optimizing more parameters in VMC always improves the nodes of the trial wave functions and therefore lowers the DMC energy. The fact that it is in practice the case must mean that our trial wave functions are reasonably accurate. Considering now also the optimization of the orbitals, it seems that the gain in the DMC total energy with respect to the situation when only the Jastrow factor and the CI coefficients are optimized is quite small if there is any. More precisely, the DMC energies obtained using the two sets of variational parameters are almost always compatible within one, two, or three standard deviations, even for small numbers of determinants. Reoptimizing the orbitals has thus only a small effect on the nodes of these trial wave functions, at least for the systems considered here.

The VMC and DMC total energies obtained at the three levels of optimization using the largest CI expansions are reported in Table~\ref{final_energies}. This table shows that the VMC and DMC errors with respect to the estimated exact energies~\cite{davidson} are reduced by about a factor of 2 when going from the optimization of the Jastrow factor only to the optimization of the Jastrow factor and the CI coefficients. As already noticed on Figures~\ref{conv_neutral_vmc} and ~\ref{conv_neutral_dmc}, the effect of reoptimizing the orbitals is very small. The best DMC total energies obtained in the present work are much lower than the DMC total energies obtained with fully reoptimized Jastrow-full-valence-complete-active-space wave functions~\cite{TouUmr-JCP-08}. This shows the importance of including excited determinants beyond the valence orbital space for improving the nodes of the wave function. We note that lower DMC energies have been reported with truncated CI wave functions for some atoms in the series in the literature~\cite{MorMcmClaKimScu-JCTC-12,canadian}. The remaining errors in the present DMC energies are essentially due to the limited basis set used and to the exclusion of core excitations in the CI expansions. The impact on the DMC energy of core excitations together with the effect of using an appropriate basis set for core correlation was illustrated in a previous work by Giner~\textit{et al.}~\cite{canadian} and will not be repeated here.

\begin{figure*}
\includegraphics[scale=0.30,angle=-90]{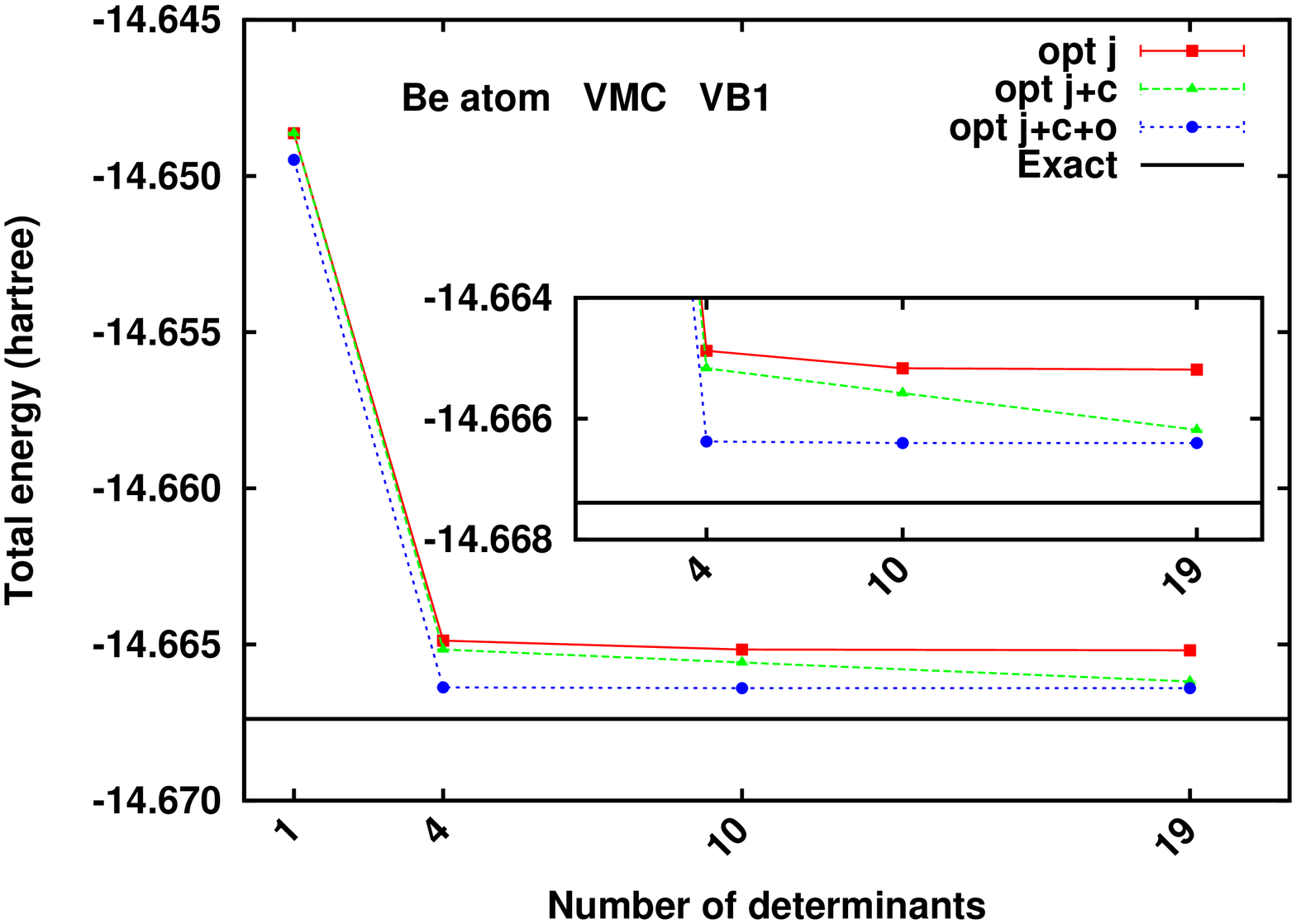}
\includegraphics[scale=0.30,angle=-90]{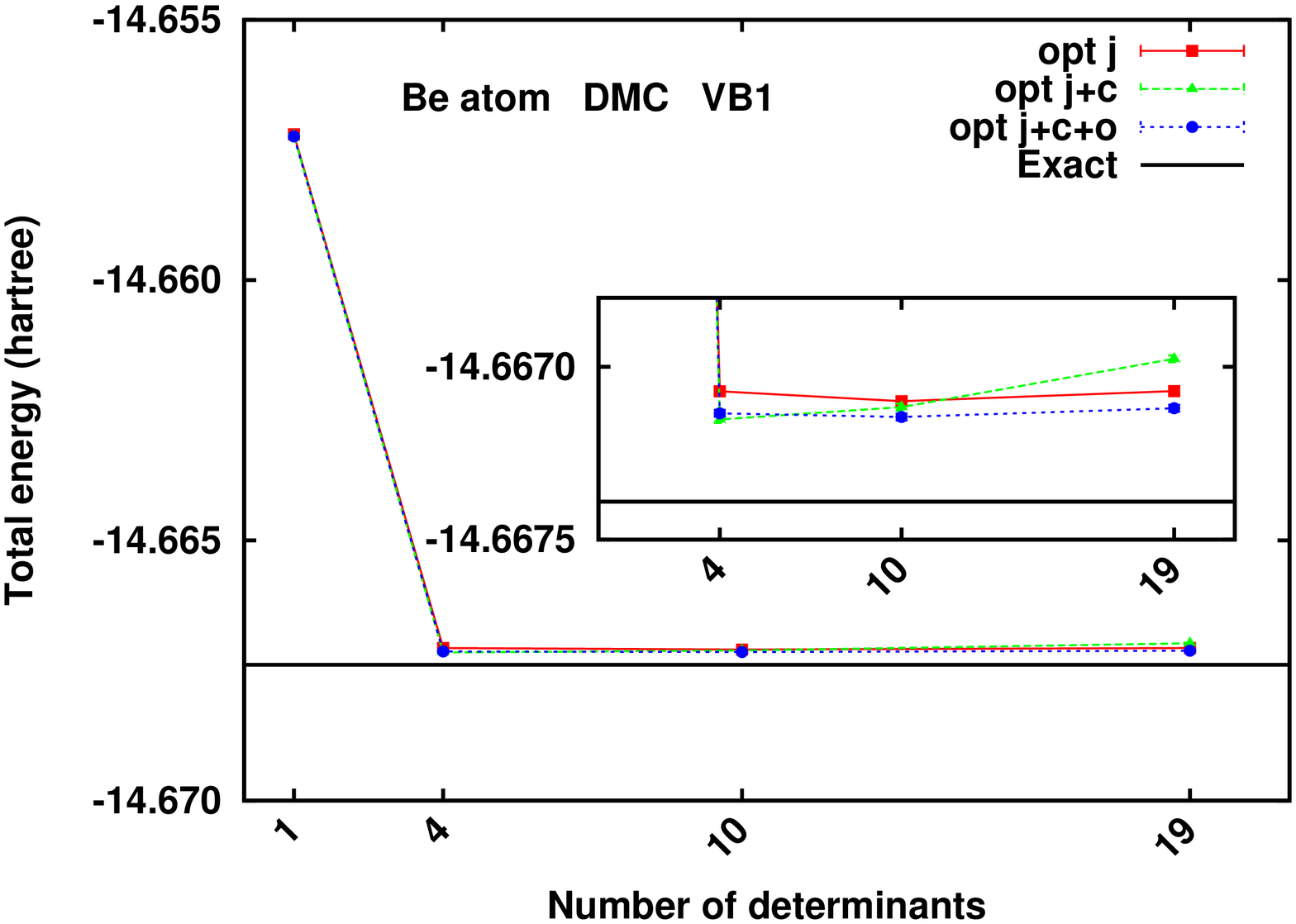}
\caption{VMC and DMC total energies of the Be atom with Jastrow-CIPSI wave functions with increasing numbers of determinants and different levels of optimization in VMC: optimization of the Jastrow factor (opt j), optimization of the Jastrow factor and the coefficients of the determinants (opt j+c), and optimization of the Jastrow factor, the coefficients of the determinants, and the orbitals (opt j+c+o). The error bars are smaller than the point symbols. The basis set used is VB1.
}
\label{be_vb1}
\end{figure*}
\begin{figure*}
\includegraphics[scale=0.30,angle=-90]{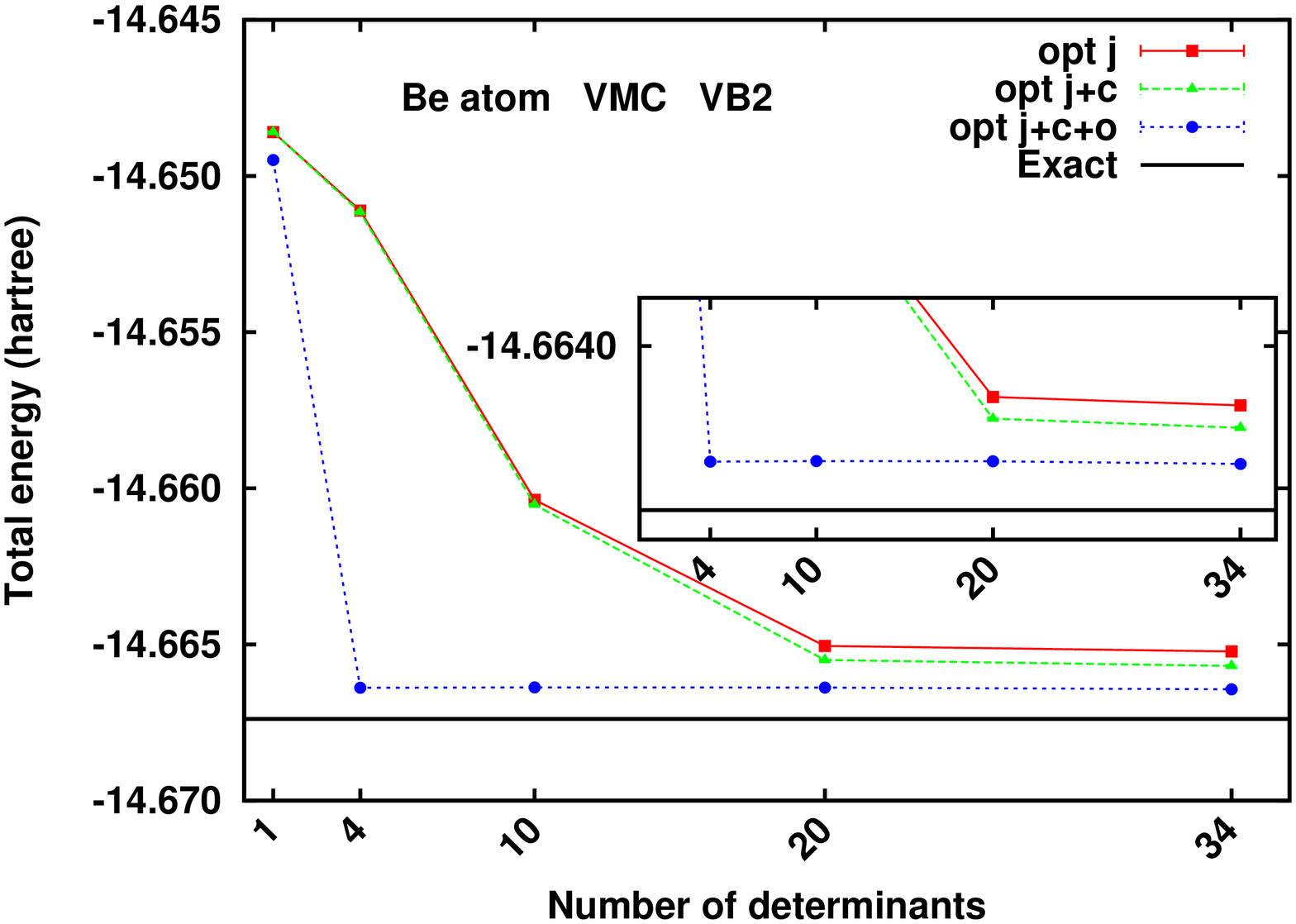}
\includegraphics[scale=0.30,angle=-90]{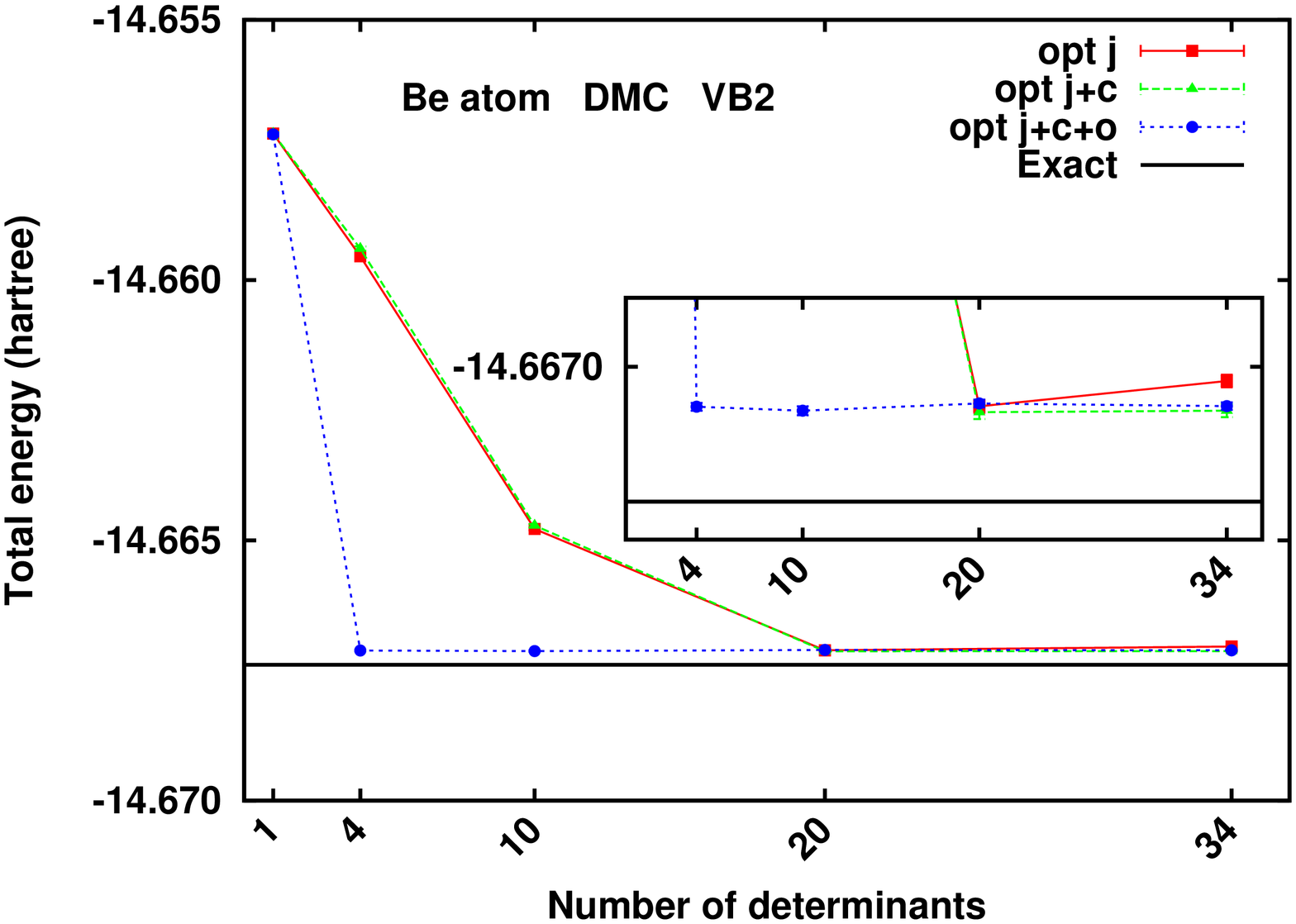}
\caption{VMC and DMC total energies of the Be atom with Jastrow-CIPSI wave functions with increasing numbers of determinants and different levels of optimization in VMC: optimization of the Jastrow factor (opt j), optimization of the Jastrow factor and the coefficients of the determinants (opt j+c), and optimization of the Jastrow factor, the coefficients of the determinants, and the orbitals (opt j+c+o). The error bars are smaller than the point symbols. The basis set used is VB2. 
}
\label{be_vb2}
\end{figure*}

\begin{table*}[t]
\caption{Total VMC and DMC energies of the Be atom with Jastrow-CIPSI wave functions with different numbers of determinants ($N_\text{det}$) and different levels of optimization in VMC: optimization of the Jastrow factor (opt j), optimization of the Jastrow factor and the coefficients of the determinants (opt j+c), and optimization of the Jastrow factor, the coefficients of the determinants, and the orbitals (opt j+c+o). The basis sets used are VB1 and VB2.}
\label{be_energies}
\begin{tabular}{l l c c c c c c c c c}
\hline\hline
        &       & \multicolumn{3}{c}{$E_\text{VMC}$} &&& \multicolumn{3}{c}{$E_\text{DMC}$} &  $E_\text{exact}^a$\\
                \cline{3-5} \cline{7-10} 
Basis set   & $N_\text{det}           $  & opt j          & opt j+c          & opt j+c+o        &&& opt j           & opt j+c         & opt j+c+o      &   \\
\hline
VB1  & 4               & -14.66488(1)  & -14.66517(1)  &  -14.666379(5) &&& -14.66707(1) & -14.66715(1) & -14.66713(1) &   \\
     & 19              & -14.665190(5) & -14.666186(5) &  -14.666402(5) &&& -14.66707(1) & -14.66698(1) & -14.66712(1) &            \\
\\
VB2  & 4               & -14.65111(3)  & -14.65116(3) &  -14.66639(3) &&& -14.65954(4) & -14.65939(4) & -14.66712(1) &   \\
     & 34              & -14.66522(3)  & -14.66569(3) &  -14.66643(3) &&& -14.66704(2) & -14.66713(2) & -14.66711(1) &  -14.66739\\
\hline\hline
\multicolumn{11}{l}{$^a$Estimated non-relativistic total energies from Ref.~\onlinecite{davidson}.}
\end{tabular}
\end{table*}

\subsection{The case of the Be atom}
\label{results_be}

Now we investigate in more details the effect of the basis set and of the level of optimization in the case of the Be atom which is known to present important static correlation effects due to the near degeneracy of the $2$s and $2$p shells. For this system, we use the VB1 and VB2 basis sets~\cite{EmaGarRamLopFerMeiPal-JCC-03} and the largest CIPSI wave functions correspond to the FCI limit within these basis sets with the $1$s frozen-core approximation. The minimal multideterminant wave function contains four determinants, the RHF determinant and the three double excitations $2$s$ \; \rightarrow 2$p$_i$ (with $i=x,y,z$). These three excited determinants are responsible for the strong multideterminant character of the ground-state wave function. We report in Figures \ref{be_vb1} and \ref{be_vb2} the convergence of the VMC and DMC total energies with the number of determinants for the VB1 and VB2 basis sets, respectively. Some corresponding energies are also given in Table~\ref{be_energies}.

With the VB1 basis set, the results are globally similar to the results obtained for the atoms from B to Ne. The VMC and DMC energies are almost converged using the four-determinant wave function. In comparison to the situation where only the Jastrow factor and the coefficients of the determinants are optimized, the convergence of the VMC energy with the number of determinants is a bit faster when the orbitals are reoptimized in VMC. As regards the DMC calculations, we observe that the DMC energy slightly increases when going from 4 to 19 determinants when optimizing only the Jastrow factor and the coefficients of the determinants. This effect can be seen thanks to the smallness of the error bars, and we believe that it is real, i.e. not due to a failure of the optimization but simply due to the fact that minimizing the VMC energy does not necessarily lead to a lower DMC energy. This slight non-monotonic behavior of the DMC energy is eliminated when the orbitals are also optimized.

With the VB2 basis set, when the orbitals are not reoptimized, the convergence of the VMC and DMC energies with the number of determinants is much slower than with the VB1 basis set. Also, the reoptimization of the coefficients of determinants in the presence of the Jastrow factor does not lead to any significant lowering of the VMC and DMC energies. Focussing on the four-determinant wave function, the DMC energy obtained when only the Jastrow factor and the coefficients of the determinants are optimized is 7.8 mhartree higher with the VB2 basis set with respect to the value obtained at the same level of optimization with the VB1 basis set (see Table~\ref{be_energies}). The reoptimization of the orbitals in the four-determinant wave function has a very large impact on both the VMC and DMC energies, and allows one to avoid this deterioration effect upon increasing the size of the basis set.

One may wonder about the origin of this deterioration observed with the RHF orbitals when going from the VB1 to the VB2 basis set, since one would naively expect an improvement when increasing the number of basis functions. This effect is related to the much more diffuse character of the RHF 2p orbitals obtained with the VB2 basis set compared to those obtained with the VB1 basis set. Indeed, in accordance to Koopmans' theorem, a RHF virtual 2p orbital of the neutral Be atom represents an approximation of an occupied 2p orbital of the Be$^-$ anion, which is very diffuse. The VB1 basis set contains only one compact 2p basis function and thus does not have any flexibility to create such a diffuse 2p orbital. By contrast, the VB2 basis set also contains a diffuse 2p basis function, and therefore the RHF optimization procedure has the flexibility to generate a much more diffuse 2p orbital. Even though such diffuse 2p orbitals are better approximations to the exact virtual RHF 2p orbitals, they are much worse to describe ground-state electronic correlation when used in multideterminant expansions. The reoptimization in VMC of the orbitals in the four-determinant wave function with the VB2 basis set leads to much more compact 2p orbitals which better describe ground-state correlation. This explanation is confirmed by the calculation of the expectation value of $r^2$ over one 2p orbital. With the VB2 basis set, this quantity is equal to $28.3$ bohr$^2$ at the RHF level, and decreases to $7.7$ bohr$^2$ after reoptimization of the orbitals in VMC using a four-determinant wave function. With the VB1 basis set, the same quantity is always equal to $7.7$ bohr$^2$. This highlights the importance of orbital optimization in cases with important static correlation effects.

Finally, we note that, with the VB2 basis set and with the largest number of determinants, the VMC energy obtained when optimizing all the parameters is significantly lower than the one obtained when optimizing only the Jastrow factor and the coefficients of the determinants. This results shows the impact of the reoptimization of the 1s orbital in the presence of the Jastrow factor. This effect is not accounted for in the FCI wave function with the frozen-core approximation. However, the reoptimization of the 1s orbital has no effect on the DMC energy.

\section{Conclusions}
\label{conclusions}

In this work, we have explored the use in VMC and DMC of trial wave functions consisting of a Jastrow factor multiplied by a truncated CI expansion in Slater determinants obtained from a prior CIPSI calculation. In the CIPSI algorithm, the CI expansion is iteratively enlarged by selecting the best determinants using perturbation theory, which provides an optimal and automatic way of constructing truncated CI expansions approaching the FCI limit and not based on \textit{a priori} criteria on orbital active spaces and/or excitation classes. 

All-electron QMC calculations on a series of atoms ranging from B to Ne with the Slater VB1 basis set and different levels of optimization of the parameters (Jastrow parameters, coefficients of the determinants, and orbital parameters) show that:\\
(1) If the coefficients of the determinants are not reoptimized in VMC in the presence of the Jastrow factor (i.e., kept as they have been obtained in a large CIPSI calculation), the VMC and DMC total energies do not always systematically decrease monotonically with respect to the number of determinants.\\
(2) If the coefficients of the determinants are reoptimized in VMC simultaneously with the Jastrow factor, an important energetic gain is obtained in VMC and DMC (the errors in the total energies are reduced by about a factor of 2), even for large numbers of determinants, and both VMC and DMC total energies converge nearly monotonically with respect to the number of determinants.\\
(3) The reoptimization in VMC of the orbitals, together with the Jastrow factor and the coefficients of the determinants, has a much smaller effect on the VMC and DMC total energies.\\
In addition, the more detailed study of the Be atom, representing a prototype of a system with important static correlation effects, shows that in this case the reoptimization in VMC of the orbitals can have a large impact on both the VMC and DMC energies when using the larger Slater VB2 basis set.

Thus, even though these conclusions should be checked on more systems, Jastrow-CIPSI wave functions with reoptimized coefficients of the determinants (and, in some cases, with reoptimized orbitals) appear as promising, systematically improvable trial wave functions for QMC calculations. Future possible works on this topic include checking the accuracy of energy differences obtained from reoptimized Jastrow-CIPSI wave functions, and performing the selection of the best determinants in the truncated CI expansions directly in QMC. The exploration of this last strategy would be indeed interesting since one can expect that the dynamic correlation brought by the Jastrow factor impacts the selection.

\section*{Acknowledgements}

It is a great pleasure to dedicate this paper to Andreas Savin who is among the early contributors to the development of quantum Monte Carlo for quantum chemistry~\cite{FlaSavPre-JCP-92,FlaSavScuNicPre-CPL-94,FlaSav-JCP-95,FlaCafSav-INC-97} and who has always encouraged this line of research. We thank Michel Caffarel, Anthony Scemama, and Cyrus Umrigar for discussions.

\end{document}